# Distance dependence of force and dissipation in non-contact atomic force microscopy on Cu(100) and Al(111)

O Pfeiffer, L Nony, R Bennewitz, A Baratoff and E Meyer

Institute of Physics, University of Basel, 4056 Basel, Switzerland

**Abstract**
The dynamic characteristics of a tip oscillating in the nc-AFM mode in close vicinity to a Cu(100)-surface are investigated by means of phase variation experiments in the constant amplitude mode. The change of the quality factor upon approaching the surface deduced from both frequency shift and excitation versus phase curves yield to consistent values. The optimum phase is found to be independent of distance. The dependence of the quality factor on distance is related to 'true' damping, because artefacts related to phase misadjustment can be excluded. The experimental results, as well as on-resonance measurements at different bias voltages on an Al(111) surface, are compared to Joule dissipation and to a model of dissipation in which long-range forces lead to viscoelastic deformations.

## 1. Introduction

Non-contact atomic force microscopy (nc-AFM) is a fascinating tool to measure small forces, beyond its ability to produce images with a high spatial resolution. It has been shown that different forces, such as long-range electrostatic, van der Waals and short-range forces, could be detected on a wide variety of surfaces [1–5]. However, extracting such quantities from the measured quantities still remains difficult, essentially because of three reasons. First, the force measurement is implicit, i.e. the experiments record the shifted resonance frequency of the lever induced by the interaction force, but not to the force itself. Secondly, as the tip is the probe and the source of the interaction, an accurate description of its shape is thus required to explain any quantitative value of the force. Thirdly, the distance between the tip and the surface is not exactly known; this introduces a further parameter for the interpretation of the data. Overcoming these difficulties has remained the purpose of many studies dealing with this subject. Dürig [6] and Giessibl [7] proposed mathematical methods to convert frequency versus distance measurements into force versus distance curves. These methods are useful because the only assumption is that the force vanishes beyond the furthest data point. Alternatively, the extracted force curves can be fitted with analytic models and quantitative values for physical variables can be extracted, but the third difficulty remains. Although these methods can in principle be extended to dissipative forces, only a few publications [8–11] have dealt with the measurement of such forces. In contrast to the rather well understood frequency shift due to conservative forces, the origin of the observed distance-dependent damping of the cantilever oscillation during nc-AFM experiments is not fully explained and still attracts much interest [12].

This work reports an experimental study of forces arising on Cu(100) and Al(111) samples during nc-AFM experiments in ultrahigh vacuum (UHV). In order to measure conservative and dissipative forces, frequency shift ($\Delta f$) and cantilever excitation ($A_{\text{exc}}$) versus distance are usually recorded, while keeping the tip oscillation $A$ constant and the phase of the excitation on resonance. But to gain more information about the whole oscillator dynamics another type of experiment, where the frequency of the cantilever at a given distance from the sample is swept, is sometimes performed, while the changes in the oscillation amplitude are recorded. For that purpose, all feedback controllers are switched off and a constant excitation amplitude is used. Such experiments reveal bifurcation patterns of the amplitude and of the phase which can occur even in non-contact range due to the nonlinear behaviour of the oscillator in interaction with the surface [13–15]. Extracting quantitative values from such experiments is not straightforward and requires modelling the nonlinear behaviour of the cantilever, as shown most recently by Polesel-Maris *et al* [16]. From an experimental point of view, performing such an experiment in UHV is not trivial. Indeed, the quality factor ($Q$) of the cantilever in UHV is high ($\sim$30 000), which means that the time constant of the cantilever ($f_0 \cong 150$ kHz) is long ($\cong$400 ms). Therefore, the frequency must be swept slowly while the drift of the piezoelectric ceramic sample holder has to be negligible.

To perform analogical measurements, we prefer instead an alternative method, referred to as phase-variation [17]. For that purpose, the distance controller is disabled while the amplitude controller remains active. Initially, the tip is at a given mean distance from the surface, and the phase between excitation and oscillation is varied. The frequency and the excitation amplitude required to maintain constant amplitude are acquired. The distance is then changed and another phase variation curve is recorded. There are two main advantages of this technique compared to sweeping the frequency: first, the large time constant for changes in the amplitude due to the high $Q$-factor is overcome, as originally mentioned by Albrecht *et al* [18] and therefore the phase sweep can be performed quicker because the amplitude regulator responds faster. Acquiring one curve takes 3 s, reducing effects of drift and creep. Secondly, the trajectory of the tip with respect to the surface remains the same. As a result, the quantities $\tilde{F}_c$ and $\tilde{F}_d$ which enter the analytic equations below remain constant. This makes them easier to apply than in the frequency sweep method (see section 2.1). Phase variation curves, where the amplitude is kept constant, can also help to interpret damping images.

## 2. Analytic description of nc-AFM

This section summarizes the description of nc-AFM by modified equations for a harmonic oscillator in interaction with a surface. Several authors have developed perturbation methods [19–22], while others applied a variational method based on the principle of least action [23–25] to derive analytical expressions which relate the frequency shift and excitation amplitude to the force versus distance, in the limit where the force is much smaller than the restoring force of the cantilever.

These expressions are compared to our experimental results in section 4.1. We first consider conservative forces without any additional dissipation. Then a local dissipation is introduced, described by a viscoelastic response to the time-dependent force. A second mechanism, Joule dissipation, is considered as well.

### 2.1. Basic equations

The observed waveform of the tip deflection naturally leads to the following trial function for its instantaneous displacement with respect to its static equilibrium position:

$$z(t) = A\cos(\omega t + \varphi). \quad (1)$$

The above-mentioned calculations then yield two coupled equations which determine the steady state of the oscillator, namely its amplitude and phase, $A$ and $\varphi$, respectively:

$$\cos(\varphi) = \frac{A}{A_{\text{exc}}}\left(1 - \frac{f^2}{f_0^2}\right) + \frac{2}{k_c A_{\text{exc}}}\tilde{F}_c \quad (2)$$

$$\sin(\varphi) = -\frac{f}{f_0}\frac{A}{A_{\text{exc}}}\left(\frac{1}{Q_0} + \frac{2}{k_c A}\tilde{F}_d\right) = -\frac{f}{f_0}\frac{A}{A_{\text{exc}} Q_{\text{eff}}}, \quad (3)$$

where $f_0$ is the fundamental resonance frequency and $Q_0$ the $Q$-factor of the free cantilever, $k_c$ its bending spring constant, $A_{\text{exc}}$ the excitation amplitude and $f$ is the frequency. By definition $Q_0 = (2\pi f_0)/\Gamma_0$, where $\Gamma_0$ is also referred to as the intrinsic damping rate of the free cantilever. Equations (2) and (3) are formally those of a driven harmonic oscillator, except that the frequency and width of the resonance are amplitude and distance dependent.

$Q_{\text{eff}}$ is an effective $Q$-value when dissipation occurs. $\tilde{F}_c$ and $\tilde{F}_d$ are the first Fourier components of the conservative and dissipative forces experienced by the tip during one oscillation period $T$, respectively (see equations (6) and (7) and (13) below). As long as $F_c \ll k_c A$, the resonance condition (maximum $A$ at fixed $A_{\text{exc}}$ or minimum $A_{\text{exc}}$ at fixed $A$) is achieved at $\varphi = -90°$ for a frequency shift $\Delta f = f - f_0 \ll f_0$. Moreover, if $F_d \ll k_c A$, the resonance width remains small compared to $f_0$, but not necessarily compared to $\Delta f$. Equations (2) and (3) can be converted into the following set of equations:

$$A_{\text{exc}} = A\sqrt{\underbrace{\left[1 - \left(\frac{f}{f_0}\right)^2 + \frac{2}{k_c A}\tilde{F}_c\right]^2}_{(a)} + \underbrace{\left[\frac{f}{f_0 Q_{\text{eff}}}\right]^2}_{(b)}} \quad (4)$$

$$\tan(\varphi) = \frac{-(f/f_0)Q_{\text{eff}}}{1 - (f^2/f_0^2) + \frac{2}{k_c A}\tilde{F}_c}. \quad (5)$$

### 2.2. Conservative forces

The conservative force is the gradient of the total interaction potential at the instantaneous position of the tip. It consists of different contributions, each having a characteristic dependence on the distance to the surface. In this work we found it sufficient to model the interaction as a sum of a van der Waals and an electrostatic long-range contribution because the tips used were rather blunt: $\tilde{F}_c^{\text{tot}} = \tilde{F}_c^{\text{VdW}} + \tilde{F}_c^{\text{el}}$, where each force $\tilde{F}_c$ can be calculated from $\tilde{F}_c = f \int_0^{1/f} F_c(z(t))\cos(2\pi ft + \varphi)\,dt$.

For a sphere-plane geometry the van der Waals contribution is [19, 24]

$$\tilde{F}_c^{\text{VdW}} = \frac{-HRA}{6(D^2 - A^2)^{3/2}}, \quad (6)$$

$D$ being the distance between the surface and the equilibrium position of the cantilever, $H$ and $R$ the Hamaker constant and the tip radius, respectively.

For the same model the electrostatic contribution, assuming $R \gg (D - A)$, can be approximated [19, 32]:

$$\tilde{F}_c^{\text{el}} = \frac{-\pi\varepsilon_0 R U_{\text{ts}}^2}{\sqrt{2A}\sqrt{D - A}}, \quad (7)$$

where $U_{\text{ts}}$ is the applied bias-voltage, $\varepsilon_0$ the permittivity of free space and $D - A$ is the tip–sample distance at the lower turning point of the oscillation.

## 2.3. Dissipative forces

Even in the non-contact range, energy can be dissipated owing to the non-instantaneous response of the sample or the tip to the forces or electric fields acting between them [12]. An additional damping, $\Gamma_d$, is then added to the intrinsic damping of the oscillator $\Gamma_0$. Consequently the amplitude drops and the controller has to increase the excitation to maintain the amplitude constant. To describe this effect, it is useful to introduce an effective damping and quality factor of the oscillator. Their expressions can be read from equation (3):

$$\frac{1}{Q_{\text{eff}}} = \frac{1}{Q_0} + \frac{2}{k_c A}\tilde{F}_d = \frac{\Gamma_0 + \Gamma_d}{2\pi f} = \frac{\Gamma_{\text{eff}}}{2\pi f}. \quad (8)$$

If dissipation occurs, an increase of $\Gamma_{\text{eff}}$ and a decrease of $Q_{\text{eff}}$ is therefore expected. Because measured frequency shifts are usually very small compared to $f_0$, typically $<10^{-3} f_0$, $f$ can be approximated by $f_0$ in the former expression (8) (but not in equation (2)!).

In this work, we focus our attention on two different dissipation channels. First, dissipation can arise if the sample (or the tip) is mechanically deformed due to the action of the tip. In the model introduced by Dürig [23] and expanded by Boisgard *et al* [26, 27] the sample is assumed to behave locally like a viscoelastic medium and is described by a reactive spring ($k_s$) and a resistive dashpot ($\gamma_s$) in parallel. A pure van der Waals interaction couples the oscillating tip to the viscoelastic surface. The two quantities $k_s$ and $\gamma_s$ define the characteristic time, $\tau_s = \gamma_s/k_s$ of the viscoelastic response. Depending on the value of $\tau_s$ with respect to the characteristic times of the oscillator, namely its period $T$ and its residence time $\tau_{\text{res}}$ at the closest turning point, $\tau_{\text{res}} \simeq \frac{T}{\pi}\sqrt{(2\Delta/A)}$ with $\Delta \lesssim D - A$, two limiting regimes of dissipation can be distinguished.

- Short relaxation times: $\tau_s \ll \tau_{\text{res}}$

$$\frac{Q_0}{Q_{\text{eff}}} - 1 = \frac{Q_0 \Gamma_{d,\text{short}}}{2\pi f_0}, \quad \text{with}$$

$$\Gamma_{d,\text{short}}(\Delta, A) \simeq \frac{\tau_s (HR)^2}{k_s}\frac{\pi^2 f_0^2}{108\sqrt{2}k_c}\frac{1}{\Delta^{9/2} A^{3/2}} \quad (9)$$

- Long relaxation times: $\tau_s \gg T$

$$\frac{Q_0}{Q_{\text{eff}}} - 1 = \frac{Q_0 \Gamma_{d,\text{long}}}{2\pi f_0}, \quad \text{with}$$

$$\Gamma_{d,\text{long}}(\Delta, A) \simeq \frac{(HR)^2}{\gamma_s}\frac{1}{72k_c}\frac{1}{\Delta^{7/2} A^{5/2}}. \quad (10)$$

In the limit $\tau_s \to 0$ the sample becomes purely elastic, i.e. the surface deformation follows the tip motion instantaneously without any phase delay, with the consequence that the dissipated energy goes to zero. On the other hand, for long relaxation times, the dissipated energy scales as $1/\gamma_s$, which leads to the less intuitive result that no significant additional damping should be observed for highly dissipating materials in nc-AFM. As in other models, $\Gamma_d$ is proportional to the square of the interaction strength ($HR$), whereas the amplitude and distance dependence changes in a characteristic fashion between the two regimes.

A further dissipation channel is Joule dissipation, which occurs due to the presence of a DC voltage between tip and sample. Because the tip oscillates, the resulting displacement current can at least in part be converted to resistive AC currents within the sample and the tip [28]. Roughly, most of the surface charge and bulk polarization induced by the electric field between tip and sample is located in an area $\pi s^2$, where $s$ is the larger of tip radius $R$ and the gap width $z_{\text{ts}}$. The induced transport currents spread out, leading to a spreading resistance $\Omega_s = 1/(2\pi\sigma s)$ ($\sigma$ being the electrical conductivity in each electrode). The resulting friction term in the equation of motion of the lever is then given by [28]

$$\gamma_e = U_{\text{ts}}^2 (\delta C_{\text{ts}}/\delta z_{\text{ts}})^2 (\Omega_s + \Omega_t), \quad (11)$$

$U_{\text{ts}}$ and $C_{\text{ts}}$ being the voltage and capacitance between tip and sample. For a spherical tip under the assumption $R \gg z_{\text{ts}}$,

$$\delta C_{\text{ts}}/\delta z_{\text{ts}} \approx 2\pi\varepsilon_0 R/z_{\text{ts}}. \quad (12)$$

The corresponding term in equation (3) is given by [20, 23, 21, 29]

$$\tilde{F}_d = \frac{f}{A}\int_0^{1/f} -\gamma(z)\dot{z}\sin(2\pi ft + \varphi)\, dt. \quad (13)$$

In a rough approximation consistent with our assumption $R \gg D - A$ we expect $s \sim R$, i.e. $\Omega$ to be independent of $D$ and $A$, and obtain with the help of equation (8)

$$\frac{Q_0}{Q_{\text{eff}}} - 1 = \beta\frac{1}{\sqrt{A}}\frac{1}{(D-A)^{3/2}}, \quad (14)$$

where $\beta = 8\pi^2 Q_0 U_{\text{ts}}^2 \varepsilon_0^2 R^2 f \Omega/(\sqrt{2} k_c)$. As in the model for the viscoelastic surface, the dissipation depends on $R^2$, whereas the conservative electrostatic term is proportional to $R$. As our samples (Cu, Al) are good conductors, one expects that Joule dissipation mainly occurs within the tip-material (n-doped silicon, eventually oxidized and covered by sample material). This makes it difficult to estimate $\Omega \cong \Omega_t$.

## 3. Experimental set-up

The force microscope used in this study is a home-built instrument using the dynamic mode for force detection as described in [30]. We use silicon cantilevers (Nanosensors™) with typical resonance frequencies of $f_0 = 160$ kHz and a spring constant $k_c = 30$ N m$^{-1}$. The tip is oscillating at a constant amplitude $A$ between 5 and 30 nm. All nc-AFM measurements are performed in UHV and at room temperature. In order to obtain a well conducting tip, we intentionally crashed the Si tip into the Cu sample until a stable time-averaged tunnelling current could be measured. The tip is most probably covered by Cu after this procedure.

Our nc-AFM set-up controls the phase $\varphi$ between excitation and oscillation of the cantilever using a phase-locked loop (PLL) [31]. The phase is set to a certain value, usually to +90°, using an electronic phase shifter so that the tip oscillates at the resonance frequency of the lever. When approaching the surface with the phase kept constant, the resonance frequency is shifted. To acquire an image, the tip is scanned above the sample and the distance to the surface is adjusted by means of a proportional-integral (PI) controller which keeps a given frequency shift $\Delta f$ constant. A further PI controller maintains a constant oscillation amplitude. The former one will be referred to as the distance controller and the latter one as the amplitude controller. In order to measure conservative and dissipative forces, $\Delta f$ and $A_{\text{exc}}$ are recorded at a chosen lateral location with the distance controller disabled. The tip is approached towards the surface until a stable mean tunnelling current of 5 pA is detected. Meanwhile, the resonance frequency and the output of the amplitude controller, namely the excitation amplitude required to maintain a constant amplitude, are acquired.

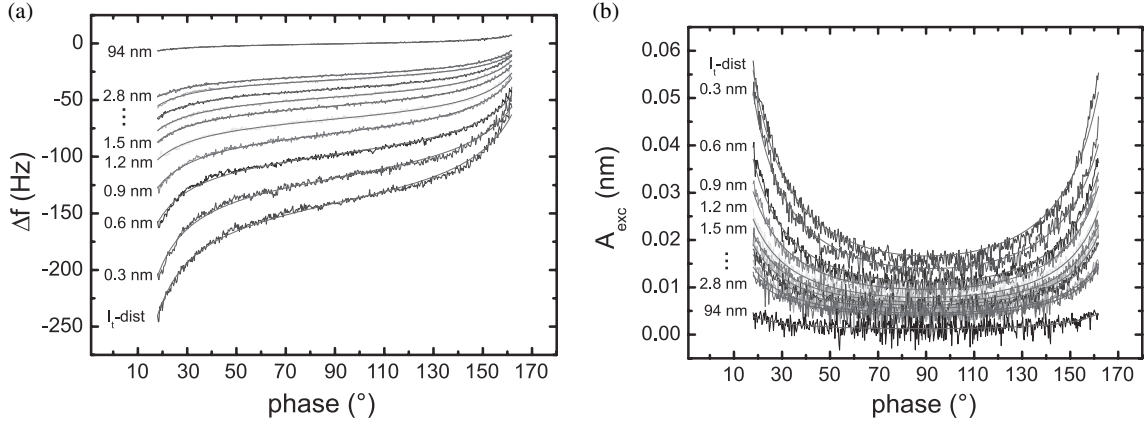

**Figure 1.** Frequency (a) and excitation amplitude (b) versus phase for different distances to the Cu(100)-sample ranging from tunnelling distance up to 94 nm. The phase chosen here is actually $\varphi + 180°$. $A = 43$ nm, $f_0 = 153\,680$ Hz, $k_c = 27$ N m$^{-1}$, $U_{\text{bias}} = 1.4$ V, $Q_0 = 34\,934$.

When performing a phase variation experiment, e.g. to check the validity of equations (4) and (5), the phase between the excitation and the oscillation of the cantilever is varied and the frequency shift and the increase of the excitation amplitude, which is needed to maintain the oscillation amplitude of the cantilever constant away from resonance, are recorded [17].

## 4. Experimental results and discussion

### 4.1. Phase-variation experiments at different mean tip–sample distances

These measurements are started in the tunnelling range; we then retracted the tip stepwise from the surface. The last pair of curves were recorded about 94 nm away from the surface. The whole set of measurements is shown in figure 1. The conservative term $\tilde{F}_c$ shifts the curves in figure 1(a) vertically, but does not change their shape. The decrease in frequency when approaching the surface can clearly be seen in figure 1(a). An appreciable decrease of the effective $Q$-value upon approaching the sample can also be deduced, because $Q_{\text{eff}}$ is in approximation inversely proportional to the slope of each curve at 90°. The increase of the slope closer to the surface can clearly be seen. Correspondingly, in figure 1(b), an increase of the excitation amplitude can be observed.

The curves 1(a) and (b) are well fitted by equations (5) and (4), respectively. The fitting parameters are $\tilde{F}_c$ and the effective $Q$-value. The spring constant $k_c$, the amplitude $A$ and the resonance frequency of the free cantilever $f_0$ are either known or calibrated quantities.

In contrast to measurements of the resonance curve with a constant excitation amplitude [14, 16], the nonlinearity of the force does not cause a distortion of the curve in phase-variation experiments. Therefore equations (4) and (5) can be checked over the full range in the constant amplitude mode because no instabilities occur, although the nonlinearity is explicitly included.

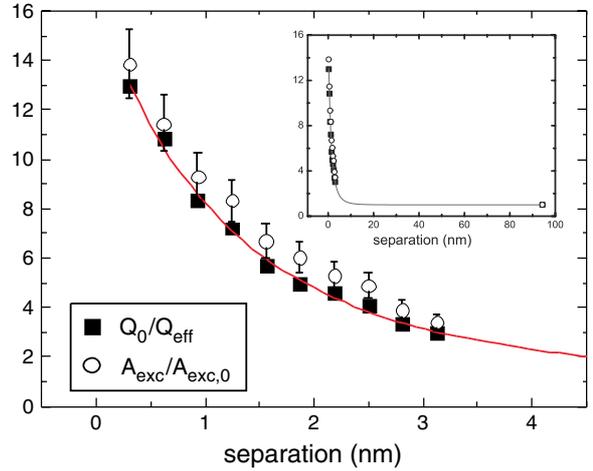

**Figure 2.** Comparison between the distance dependence of the excitation amplitude $A_{\text{exc}}$ and the effective $Q$-value deduced from figures 1(a) and (b). The main plot only shows measurements close to the sample, while the inset contains all data points. The displayed error bars of 10% for the excitation signal are mainly caused by the uncertainty of the amplitude calibration. A bias voltage of 1.4 V is applied to the sample.
(This figure is in colour only in the electronic version)

In figure 2 the distance dependences of the excitation amplitude and of the $Q$-value are compared by plotting the ratio between the excitation amplitude at a given distance ($A_{\text{exc}}$) and the excitation amplitude of the free cantilever ($A_{\text{exc},0}$), and plotting the ratio between the $Q_0$-value of the free cantilever and the effective $Q$-value, which is deduced from the fit of the curve in figure 1(a) with equation (5). Far away, the ratio is 1, by definition. Upon approaching the surface both ratios increase similarly. A difference of a few per cent is observed which is due to the uncertainty of the fit of the $Q$-value and of the calibration of the $A_{\text{exc}}$ signal. Close to resonance $A_{\text{exc}}$ is dominated by the term (b) in equation (4) because right on resonance ($\varphi = -90°$) term (a) vanishes.

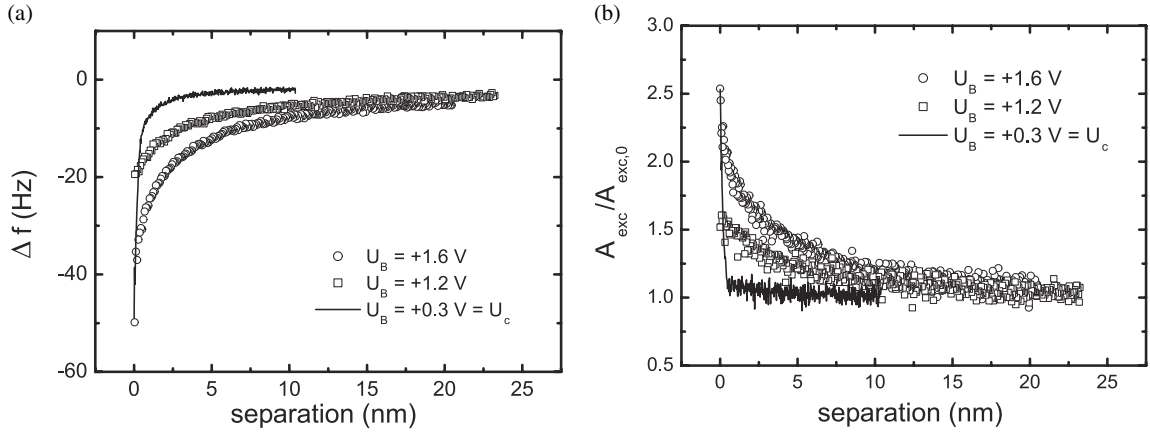

**Figure 3.** Frequency shift and ratio $A_{exc}/A_{exc,0}$ versus distance for different bias voltages recorded on Al(111). When applying +0.3 V on the sample, the contact potential is compensated and the electrostatic interaction is minimized. The electrostatic interaction adds a long-range component to both curves. $A = 6.5$ nm, $f_0 = 162\,050$ Hz, $k_c = 29.4$ N m$^{-1}$, $Q_0 = 31\,500$.

The large change in $Q$ observed in figure 2 contradicts the stability criterion proposed by Giessibl [33]. According to this criterion amplitude control can become difficult if the energy loss is large compared to the energy loss of the free cantilever. It appears that this criterion is too stringent at least for the phase variation experiments discussed here. As seen in figure 1(b) the amplitude controller properly follows the imposed phase change; indeed the $A_{exc}$ variation is satisfactorily fitted by equation (4) if $A$ is considered constant. However, attention must be paid to the fact that a measurement is performed at a constant distance from the surface and at a constant location. Amplitude control can be less perfect when scanning the surface (depending on the scan speed, the corrugation of the sample and the settings of the amplitude and distance controllers).

### 4.2. Distance dependence of frequency shift and effective Q-value

It has been demonstrated that it is possible to extract frequency shift and $Q_{eff}$ versus distance from the measured phase variation curves. A drawback of this technique is that the obtained data would only consist of 11 points and the interesting distance range (between 3 and 20 nm) is not covered. Drift effects become a problem if more than 11 phase curves are acquired. Furthermore, the considerable bias voltage applied to detect a sizable tunnelling current at closest approach complicates the analysis.

In section 4.1 we confirmed that $A_{exc}/A_{exc,0}$ is indeed equal to $Q_0/Q_{eff}$ as long as the optimum phase of +90° is maintained during an experiment. This demand is fulfilled with a properly adjusted PLL. Then, nc-AFM efficiently separates conservative and dissipative interactions. Therefore it is more efficient to simultaneously record $\Delta f$ and $A_{exc}$ versus distance curves at that phase setting to overcome the problems discussed above. Such measurements were performed separately at different bias voltages to analyse the distance dependence of frequency shift and dissipation (see figure 3). These experiments where performed on Al(111) with a different cantilever ($f_0 = 162\,050$ Hz, $k_c = 29.4$ N m$^{-1}$, $Q_0 = 31\,500$), oscillating at an amplitude of $A = 6.5$ nm.

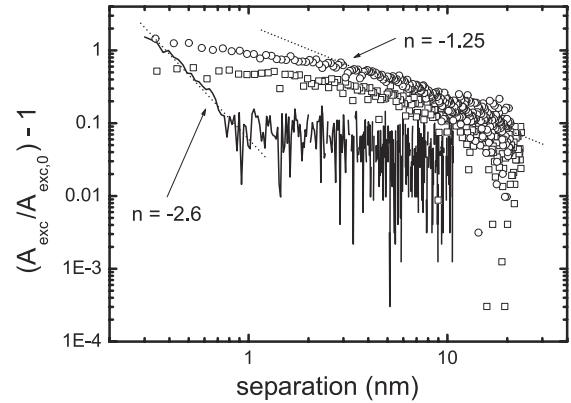

**Figure 4.** Log–log-plot of the increase of the additional excitation signal $(A_{exc}/A_{exc,0}) - 1$ measured on Al(111). The solid curve represents the measurement at compensated contact potential, the open squares are recorded at a sample bias of +1.2 V and the open circles at +1.6 V. In the last two cases the long-range part of each curve can be fitted with an exponent of about $n = -1.25$. The measurement with the compensated contact potential only allows a fit below 1 nm with $n \approx -2.6$ if the closest point is set to 0.3 nm. $A = 6.5$ nm, $f_0 = 162\,050$ Hz, $k_c = 29.4$ N m$^{-1}$, $Q_0 = 31\,500$.

The shape of the $A_{exc}$ curve measured at a bias voltage of 1.6 V looks qualitatively like that obtained during the phase variation experiment on Cu(100) (see figure 2). The long-range effects due to the applied bias voltage can clearly be seen in both frequency and $A_{exc}$ curves. The curves measured at compensated contact potential show much steeper changes at small distances. The predicted distance dependence of $\Delta f$ and of $A_{exc}$ is given by different inverse power laws $d^n$, where $n < 0$; see equations (6), (7) for $\Delta f$ and equations (9), (10), (14) for dissipation. Following [32] we attempted to extract exponents from the slope of measured curves in log–log-plots. Particular attention must be paid to the origin of the distance. The point acquired closest from the surface was assigned a minimum tip–sample separation of 0.3 nm, a typical distance where the tunnelling current becomes measurable [9].

For the measurements with an applied bias voltage (of 1.2 or 1.6 V), an exponent of about $n = -1.25$ is obtained for the long-range part of the dissipation (see figure 4). This value

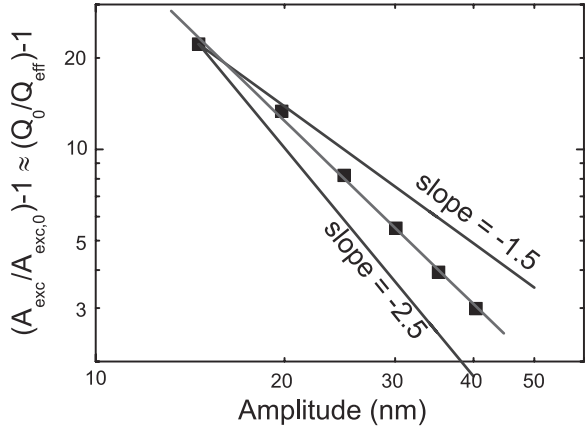

**Figure 5.** Log–log plot of the increase in the excitation amplitude for different oscillation amplitudes, when the tip comes within tunnelling range to the Cu(100) sample. The slope of the measured curve lies between the values $-1.5$ and $-2.5$, which correspond to the model for short and for long relaxation time of a viscoelastic surface. $A = 43$ nm, $f_0 = 153\,680$ Hz, $k_c = 27$ N m$^{-1}$, $U_{bias} = 1.4$ V, $Q_0 = 34\,934$

is in good agreement with $n = -1.3$ obtained by Stipe *et al* [34], even though their experimental set-up is simpler: in their case the tip oscillates parallel to the surface. This exponent $n = -1.25$ is also close to the value obtained in the simple model for the Joule dissipation $n = -1.5$ (see equation (14)). For the frequency shift an exponent of about $n = -0.6$ was found, which is in good agreement with equation (7).

For the experiment with compensated contact potential, the dissipation is buried in the noise in the long-range regime. For separations below 1 nm, an exponent of about $n = -2.6$ was found. As mentioned in the introduction, this value sensitively depends on the chosen origin of the distance axis (e.g. an additional offset of 0.3 nm changes the exponent to $n = -4.3$). This fact makes it very difficult to assign the measurements to a particular mechanism; the viscoelastic model of the surface gives a distance exponent between $-3.5$ and $-4.5$. Since the tip comes very close to the surface, it is clear that short-ranged chemical forces can play an important role and should be included in the future.

*4.3. Amplitude dependence of the effective Q-value*

The effective $Q$-value in both models for the dissipation also depends on the oscillation amplitude $A$ of the cantilever. The analytical expressions for the Joule dissipation and for the two limits of a long or a short mechanical relaxation time of the surface show three different dependences of $A$. To determine the dependence of $Q_{eff}$, the excitation amplitude $A_{exc}$ was measured in tunnelling range for different oscillation amplitudes $A$ on Cu(100) for $U_{bias} = 1.4$ V.

In figure 5, $(A_{exc}/A_{exc,0}) - 1$ is plotted. In a log–log plot, a linear dependence is in fact observed, as predicted by equations (9), (10) and (14). However, the slope of the fit is $-2$, which lies between the limiting predictions of $A^{-1.5}$ and $A^{-2.5}$ of the viscoelastic model. The $A^{-1/2}$ dependence predicted for Joule dissipation is not observed, presumably because other dissipation channels are more important at the small minimum separation chosen.

*4.4. Geometry of the tip*

All previous measurements depend on the tip geometry. If we assume a van der Waals or electrostatic tip–sample interaction the frequency shift depends linearly on the tip radius $R$, but the change in $A_{exc}$ scales as $R^2$. Thus the shape of the tip affects $A_{exc}$ and $Q_{eff}$ more strongly than the frequency shift. When repeating phase-variation experiments with a very sharp tip, no significant increase of $A_{exc}$ could be detected down to the tunnelling range. This implies that tip changes due to gentle tip crashes can toggle between quite different levels of dissipation down to tunnelling distances [35].

## 5. Conclusions and outlook

This work is an experimental study of conservative and dissipative forces arising on Cu(100) and Al(111) samples during nc-AFM experiments in UHV. The measured dependences of the frequency shift and excitation amplitude versus phase behave qualitatively and quantitatively according to theoretical predictions. The decrease in the effective $Q$-value is indeed inversely proportional to the increase of the excitation amplitude upon approaching the tip to the surface. If the phase is properly maintained at $\varphi = -90°$ it is therefore sufficient to record $A_{exc}$ to obtain information about interaction-induced dissipation. The phase variation measurements confirm that the condition of $\varphi = -90°$ is independent of the tip–sample distance although the resonance is shifted. This indicates that $A_{exc}$ images recorded under that condition are free of phase-shift artefacts. Nevertheless, apparent dissipation could still arise owing to the finite time constants of the controllers of the microscope as emphasized by Gauthier *et al* [36] and Couturier *et al* [37].

The model of Boisgard *et al* where van der Waals forces couple to the sample and induce viscoelastic deformations appears applicable to the Cu sample in the case of a blunt tip. The dependences of the damping on the amplitude and on the distance lies between the two limiting solutions for elastic and very viscous response of the sample. Joule dissipation is an important source of dissipation at large distances in the presence of an uncompensated bias voltage and the observed distance dependence for measurements is in good agreement with the model first predicted by Denk and Pohl [28].

In the future, it would be very helpful to characterize the structure of the tip end by other means (e.g. field ion microscopy) in order to achieve a better understanding of the observed tip shape effects. A heterogeneous sample would be an ideal test surface to observe changes in the distance dependence and to compare differences in local compliance. Furthermore, it would be worth examining whether the viscoelastic dissipation model can be applied to short-range forces.

## Acknowledgments

This work was supported by the Swiss National Science Foundation, the Kommission zur Förderung von Technologie und Innovation and the National Centre of Competence in Research on Nanoscale Science.